\documentclass[preprint]{revtex4-1}
\usepackage{graphicx}
\usepackage[right]{eurosym}

\begin{document}
\title{On the ``persistency" of scientific publications: \\introducing an h-index for journals}
\author{Roberto Piazza}
\affiliation{Department CMIC ``Giulio Natta'', Politecnico di
Milano, P.zza Leonardo da Vinci 32, 20133 Milano, Italy}
\email{roberto.piazza@polimi.it}%
\date{\today}

\begin{abstract}
What do we really mean by a ``good'' scientific journal? Do we care more about the short--time impact of our papers, or about the chance that they will still be read and cited on the long run? Here I show that, by regarding a journal as a ``virtual scientist'' that can be attributed a time--dependent Hirsch $h$-index, we can introduce a parameter that, arguably, better captures the ``persistency" of a scientific publication. Curiously, however, this parameter seems to  depend above all on the ``thickness'' of a journal. 

\end{abstract}

\maketitle
They are the papers that honor the journal, not the journal that honors the papers. This is a basic truth, which anyone of us should carefully keep in mind while judging the quality of a scientific work. Nevertheless, the struggle to publish significant results in high--ranking journals, either because of an understandable desire to increase their visibility, or because of the increasing academic pressure to do so, has, willy--nilly, become the norm within the scientific community.

A widespread metrics to rank scientific journals is based on the ISI Journal Impact Factor (IF), which, in spite of several criticisms raised in the recent past, surely yields a reasonable comparison of the effectiveness of a given journal in fostering a rapid dissemination of (usually) relevant scientific results. With all possible caveats, the IF is surely a parameter the publishers need to care about. Yet, as scientists, we may surely  wonder whether the short--time impact of a paper necessarily implies its \emph{persistency} over a longer period. In other words, the question is the following: which journals contain a substantial number of works are still read, appreciated, and cited \emph{well after} their publication? I am pretty sure that most colleagues would agree with my impression that these are not necessarily those journals which stand out for their impressive IF.

\begin{figure}[t]
\centering
  \includegraphics[width=0.8\columnwidth]{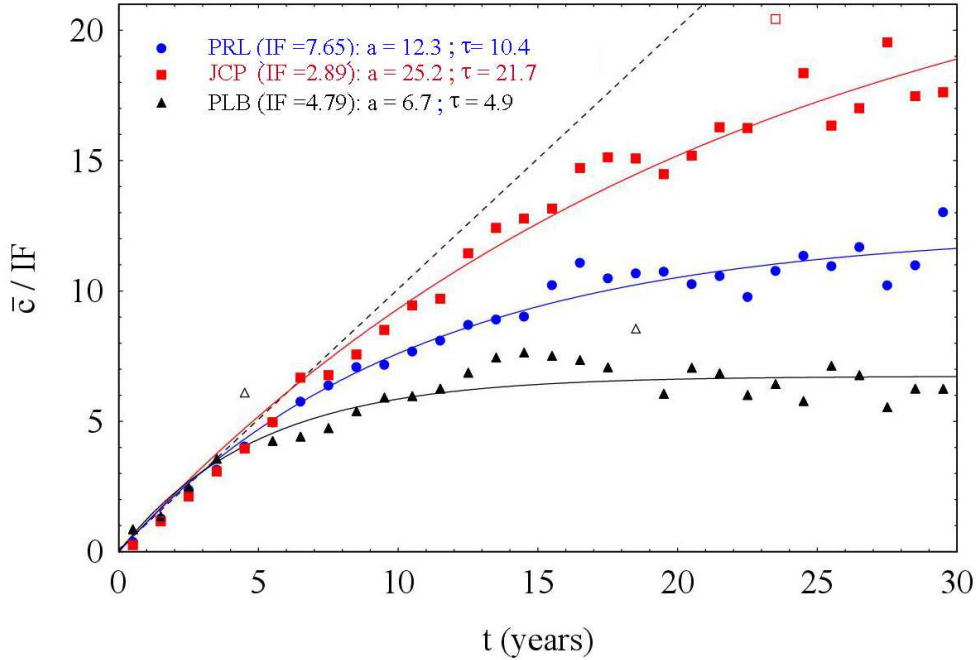}
  \caption{\label{f1}\footnotesize Average number of citations $\bar{c}$, normalized by the journal IF, as a function of time $t$ since publication for PRL (bullets), JCP (squares), and PLB (triangles). Full lines are exponential fits $c/\mathrm{IF}= a[1-\exp(-t/\tau)]$, with parameters shown in the legend. The broken line, with a unit slope, corresponds to the approximate short-time limiting behavior $\bar{c} = \mathrm{IF}t$ Open symbols refer to ``anomalous'' years that were not considered in the fit (data collected from ISI Web of Science, April 2017).}
\end{figure}

Impressions, however, must be substantiated by real data. As a physicist, let me consider just three journals in  my own field that have a rather different audience and scope: i) Physical Review Letters (PRL), a reference journal for physicists that publishes short reports covering major advances of broad interest in all aspects of physics; ii) Journal of Chemical Physics (JCP), which mostly publishes broader, circumstantial reports of significant research, expected to be of long--lasting interest for the community of chemical and condensed matter physicists; iii) Physics Letters B (PLB), which, similarly to PRL, is a letter journal, covering however only the restricted and rapidly advancing fields of high--energy physics and cosmology.  Figure~\ref{f1} contrasts the average number of citations $\bar{c}$ for all the papers\footnote{Here and in the following, by ``papers'' I mean proper articles, namely, I did not include, for instance editorial material, corrections, or review papers.}, published in one year in these journals as a function of the years since  publication, normalized, to be more meaningful, to the IF of each journal~\footnote{Note that a few ``anomalous'' years, in which the total number of citation is anomalously large because of the occurrence of one or two exceptionally cited papers, are not included in the fit. For instance, the huge number of citations for JCP in 1993 is due to a single, fundamental paper in density functional theory, which has already obtained more than 60000 citations, while the value for PLB in 2012 is mostly due to the two papers describing the discovery of the Higgs boson at CERN.},  The three data sets, which are reasonably well fitted by a saturating exponential trend, $\bar{c}/\mathrm{IF} = a[1-\exp(-t/c)]$, display a rather different long--time behavior. While for PRL, which has the highest IF, $\bar{c}$ saturates to a value which is approximately 12 times its IF, papers published in JCP asymptotically reach an average number of citation that is, in units of its IF, twice as large. Namely, on the long run an article published in JCP is typically expected to be cited only 20\% less than a letter in PRL, in spite of the fact that the ratio of the IF of these two journals is less than 0.4. As witnessed by the rather different exponential constants $\tau$, this is because many papers published in JCP continue to be read and cited well beyond a quarter of a century after their publication. Even more striking is the comparison with PRB, whose articles have a quite short ``lifetime'' and are asymptotically much \emph{less} cited than those in JPC, in spite of the higher IF.

\begin{figure}[t]
\centering
  \includegraphics[width=0.8\columnwidth]{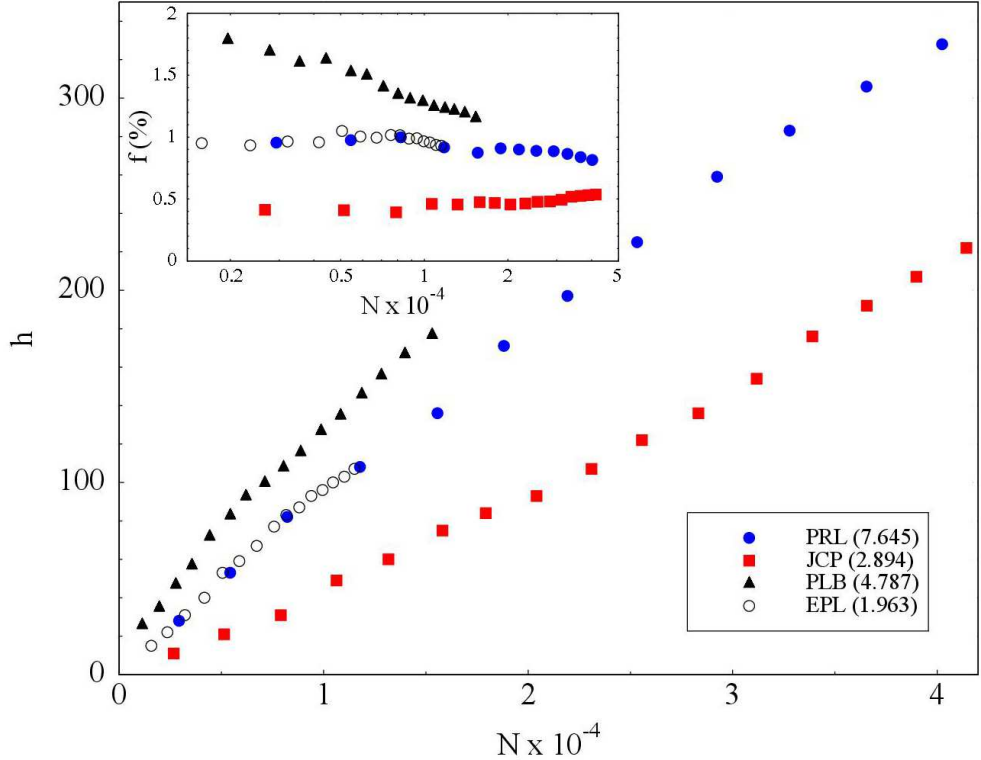}
  \caption{\label{f2} \footnotesize Body: $h$-index versus the number of published papers for the journals considered in Fig.~\ref{f1}, with the additional inclusion of Europhysics Letter/EPL, evaluated by considering all the citations obtained till April 2017 by all those papers published between 2001 and 2016 (namely, for a given journal the abscissas of the 16 data points  correspond to the total number $N$ of publications in each year between 2001 and 2016).  Inset: Fraction $f = h/N$ (per cent) of the total number of publications which have a number of citations $c\ge h$ (from ISI Web of Science).}
\end{figure}

Evidently, the IF of the journal where a paper has been published does not necessarily capture a crucial feature that makes  a scientific work truly successful, namely, the \emph{duration} of its impact on the reference community, which is what I mean by ``persistency''. We may then wonder whether an approach that better takes into account this important aspect, yielding a more robust and comprehensive ranking parameter, can be devised. As a matter of fact, a current standard evaluation parameter of the overall performance of an \emph{individual} scientist, the celebrated Hirsch $h$-index~\cite{Hirsch2005}, does exist. Indeed, with all possible caveats about its improper use, the h-index is commonly regarded at least as a first indication of the quality and persistence in time of a scientific career. Nothing prevents us, however, to try and define an $h$-index also for a scientific journal, regarded as a kind of ``virtual scientist''. On the ISI Web of Science, this is easily done by typing in the journal name, using as search field the publication name. Contributions that should not be regarded as standard scientific articles, such as editorial material, corrections, or review papers, can easily be excluded from the search by adding as a secondary  search field ``article'' as document type. The journal h-index can then be directly obtain from the Citation Report, when the number of results is less than 10000, or by a simple ranking of the latter with decreasing number of citations for larger sets of results.

The $h$-index of an individual scientist increases along his/her career: similarly, a journal $h$-index grows of course with time. Figure~\ref{f2} displays the $h$-index of four journals, obtained by considering the total number of citations obtained till April 2017 by all the papers published since the beginning of this millennium, plotted as a function of the progressively increasing total number of papers $N$ published  by each journal in the same period.  In addition to the three journals presented in Fig.~\ref{f1}, I also considered EPL (formerly Europhysics Letter), a journal letter supported by the European Physical Society (EPS), in some sense the ``European counterpart" of PRL. The figure body shows that, in a first approximation,  $h$ grows linearly with $N$, but with a slope that depends on the journal. In fact, the figure inset shows that, for PRL, JCP, and EPL, the fraction $f = h/N$ of articles belonging to the group with a number of citation $c\ge h$ is approximately constant over a period of 16 years, witnessing that $h$ is a quite stable indicator of the overall quality of the published by these journals.  A notable exception is however PLB, which shows a steady-decrease of almost a factor of two over the same period, which once again speaks of the rather limited ``lifetime" of the papers published in this journal. What is more interesting, however, is that the typical value of $f$ does not seems to be directly related to the IF of a given journal. For instance, the average value $\bar{f} = 0.98$ for EPL is very close, or even slightly higher than  that of PRL ($\bar{f}= 0.90$), in spite of the fact that the letter journal of EPS has a much lower IF than its American counterpart. Notice that, besides their IF, these two journals consistently differ for the total number of published papers in the considered period, which is in fact almost flour times larger for PRL.
\begin{figure}[t]
\centering
  \includegraphics[width=0.8\columnwidth]{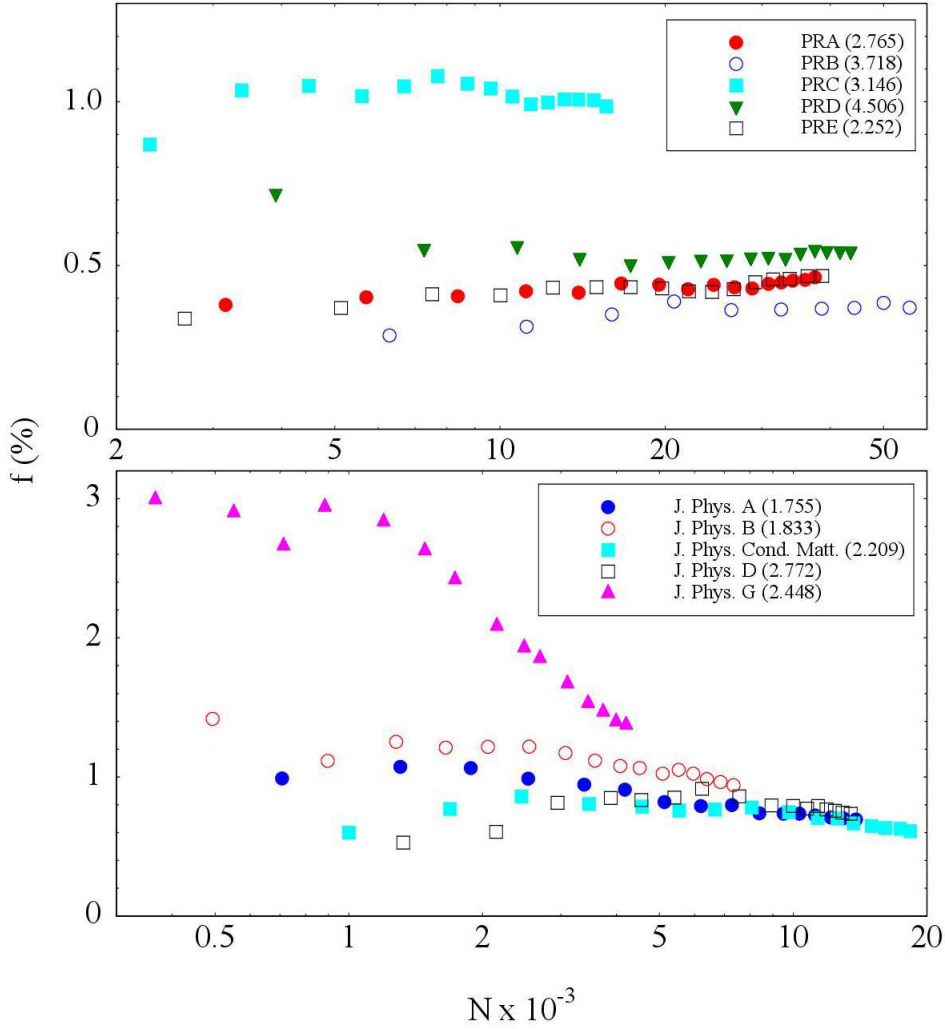}
  \caption{\label{f3}\footnotesize Upper panel: Fraction $f =h/N$ of ``highly cited" papers ($c\ge h$) versus $N$ for the Physical Review collection of the American Physical Society.  Lower panel: the same quantity for the journal of Physics collection of IOP Publishing (from ISI Web of Science).}
\end{figure}
\begin{table}[t!]
 \centering
 \caption{\label{t1}  \footnotesize Physical journals (ISO abbreviated titles) analyzed in this study, ordered by decreasing impact factor. The $h$-index of each journal is obtained by considering all the citations obtained, till April 2017, by the total number $N$ of articles published between 2001 and 2015 (corresponding to an average number of publication $n=N/16$ per year.}
 \vspace{10pt}
 \scriptsize
\begin{tabular}{|ccccccc|cccccc|} \hline
Journal & IF & $n$ & $h$& $\mathrm{IF}/h$ & $h/\sqrt{n}$ &&Journal & IF & $n$ & $h$& $\mathrm{IF}/h$ & $h/\sqrt{n}$ \\ \hline
Phys. Rev. Lett.	&	7.645	&	3320	&	413	&	1.85	&	7.17	&&	Physica E	&	 1.904	&	452	&	69	&	2.76	&	3.25	\\
J. High Energy Phys.	&	6.023	&	1294	&	170	&	3.54	&	4.73	&&	J. Synchrot. Radiat.	&	1.877	&	130	&	66	&	2.84	&	5.79	\\
Astrophys. J.	&	5.909	&	2566	&	281	&	2.10	&	5.55	&&	Mol. Phys.	&	1.837	 &	296	&	65	&	2.83	&	3.78	\\
Astrophys. J. Lett.	&	5.487	&	371	&	108	&	5.08	&	5.61	&&	J. Phys. B	&	1.833	 &	457	&	69	&	2.66	&	3.23	\\
Eur. Phys. J. C	&	4.912	&	374	&	106	&	4.63	&	5.48	&&	Int. J. Mod. Phys. A	&	 1.799	&	373	&	56	&	3.21	&	2.90	\\
Phys. Lett. B	&	4.787	&	945	&	177	&	2.70	&	5.76	&&	Physica A	&	1.785	&	 716	&	101	&	1.77	&	3.77	\\
Phys. Rev. D	&	4.506	&	2645	&	235	&	1.92	&	4.57	&&	Appl. Phys. B	&	 1.785	&	374	&	83	&	2.15	&	4.29	\\
Phys. Chem. Chem. Phys.	&	4.449	&	1406	&	150	&	2.97	&	4.00	&&	Chem. Phys.	&	 1.758	&	338	&	81	&	2.17	&	4.41	\\
Nuclear Fusion	&	4.040	&	228	&	77	&	5.25	&	5.09	&&	J. Phys. A	&	1.755	&	 877	&	95	&	1.85	&	3.21	\\
J. Mech. Phys. Solids	&	3.875	&	129	&	107	&	3.62	&	9.43	&&	Phys. Lett. A	&	 1.677	&	852	&	119	&	1.41	&	4.08	\\
Soft Matter (2005)	&	3.798	&	660	&	90	&	4.22	&	3.50	&&	Phys. Status Solidi A	 &	1.648	&	495	&	60	&	2.75	&	2.70	\\
Nucl. Phys. B	&	3.735	&	395	&	149	&	2.51	&	7.50	&&	Eur. Phys. J. E	&	1.625	 &	139	&	73	&	2.23	&	6.18	\\
Phys. Rev. B	&	3.718	&	5408	&	290	&	1.28	&	3.94	&&	Physica D	&	1.579	 &	196	&	83	&	1.90	&	5.93	\\
Biophys. J.	&	3.632	&	719	&	176	&	2.06	&	6.56	&&	J. Phys. Soc. Japan	&	1.559	 &	579	&	87	&	1.79	&	3.61	\\
New J. Phys.	&	3.570	&	530	&	119	&	3.00	&	5.17	&&	J. Stat. Phys.	&	1.537	 &	214	&	67	&	2.29	&	4.58	\\
Ann. Phys.-Berlin	&	3.443	&	54	&	35	&	9.84	&	4.76	&&	Phys. Status Solidi B	 &	1.522	&	498	&	71	&	2.14	&	3.18	\\
Astropart. Phys.	&	3.425	&	92	&	77	&	4.45	&	8.02	&&	Appl. Phys. A	&	 1.444	&	586	&	100	&	1.44	&	4.13	\\
Phys. Rev. C	&	3.146	&	959	&	153	&	2.06	&	4.94	&&	J. Biol. Phys.	&	1.394	 &	35	&	30	&	4.65	&	5.06	\\
Appl. Phys. Lett.	&	3.142	&	4473	&	293	&	1.07	&	4.38	&&	Physica B	&	 1.352	&	1040	&	77	&	1.76	&	2.39	\\
J. Chem. Phys.	&	2.894	&	2584	&	221	&	1.31	&	4.35	&&	Nucl. Phys. A	&	 1.258	&	534	&	103	&	1.22	&	4.46	\\
J. Magn. Reson.	&	2.889	&	226	&	79	&	3.66	&	5.25	&&	Eur. Phys. J. B	&	1.223	 &	395	&	79	&	1.55	&	3.97	\\
Exp. Astron.	&	2.867	&	36	&	30	&	9.56	&	4.98	&&	Eur. Phys. J. D	&	1.208	 &	290	&	63	&	1.92	&	3.70	\\
Class. Quantum Gravity	&	2.837	&	452	&	108	&	2.63	&	5.08	&&	Int. J. Mod. Phys. C	&	1.195	&	128	&	39	&	3.06	&	3.44	\\
J. Phys. D	&	2.772	&	809	&	99	&	2.80	&	3.48	&&	Phys. Scr.	&	1.194	&	524	 &	54	&	2.21	&	2.36	\\
Phys. Rev. A	&	2.765	&	2311	&	173	&	1.60	&	3.60	&&	Am. J. Phys.	&	 1.012	&	159	&	55	&	1.84	&	4.37	\\
J. Fluid Mech.	&	2.514	&	501	&	121	&	2.08	&	5.41	&&	Philos. Mag. Lett.	&	 0.918	&	89	&	44	&	2.09	&	4.66	\\
Med. Phys.	&	2.496	&	484	&	116	&	2.15	&	5.27	&&	Z. Naturfors. A	&	0.886	&	 122	&	35	&	2.53	&	3.17	\\
J. Phys. G	&	2.448	&	269	&	58	&	4.22	&	3.54	&&	Phase Transit.	&	0.858	&	 89	&	27	&	3.18	&	2.86	\\
Philos. Trans. R. Soc. A	&	2.441	&	169	&	86	&	2.84	&	6.62	&&	Int. J. Mod. Phys. B	&	0.850	&	462	&	37	&	2.30	&	1.72	\\
Eur. Phys. J. A	&	2.373	&	217	&	67	&	3.54	&	4.54	&&	Physica C	&	0.835	&	 679	&	59	&	1.42	&	2.26	\\
Phys. Rev. E	&	2.252	&	2401	&	181	&	1.24	&	3.69	&&	Tech. Phys. Lett.	&	 0.702	&	330	&	23	&	3.05	&	1.27	\\
J. Phys.-Condens. Matter	&	2.209	&	1155	&	111	&	1.99	&	3.27	&&	Eur. Phys. J. - Appl. Phys.	&	0.664	&	162	&	31	&	2.14	&	2.44	\\
J. Phys. Chem. Solids	&	2.048	&	350	&	66	&	3.10	&	3.53	&&	Tech. Phys.	&	 0.569	&	296	&	23	&	2.47	&	1.34	\\
Phys. Fluids	&	2.017	&	491	&	98	&	2.06	&	4.42	&&	Ferroelectrics	&	0.491	 &	389	&	30	&	1.64	&	1.52	\\
Europhys. Lett. / EPL	&	1.963	&	715	&	107	&	1.83	&	4.00	&&	Solid State Technol.	&	0.200	&	71	&	12	&	1.67	&	1.43	\\
\hline
\end{tabular}
\end{table}

Similar considerations can be applied to the data shown in Fig.~\ref{f3}, where we plot the dependence on $N$ of the $h$-index for the main collection of physical journals published, respectively, by the American Physical Society (the Physical Review series, where the acronym ``PRA'' stands for instance for ``Physical Review A") and by the British Institute of Physics Publishing (the Journal of Physics series). We can notice that the fraction $h/N$ is, approximately, constant in time, with the notable exception of J. Physics G, once again a journal publishing theoretical and experimental research in nuclear and particle physics. Also in this case, there is no simple relation between the fraction $f$ and the journal IF. Furthermore, we can observe that the values of $f$ for the J. Phys. journals are larger than those for the Physical Review journals, although their impact factor is typically lower. Like in the case of EPL vs. PRL, we can however notice  that the values of $N$ for the journals of the J. Phys. series are quite lower than for the Physical Review journals.

\begin{figure}[t]
\centering
  \includegraphics[width=0.8 \columnwidth]{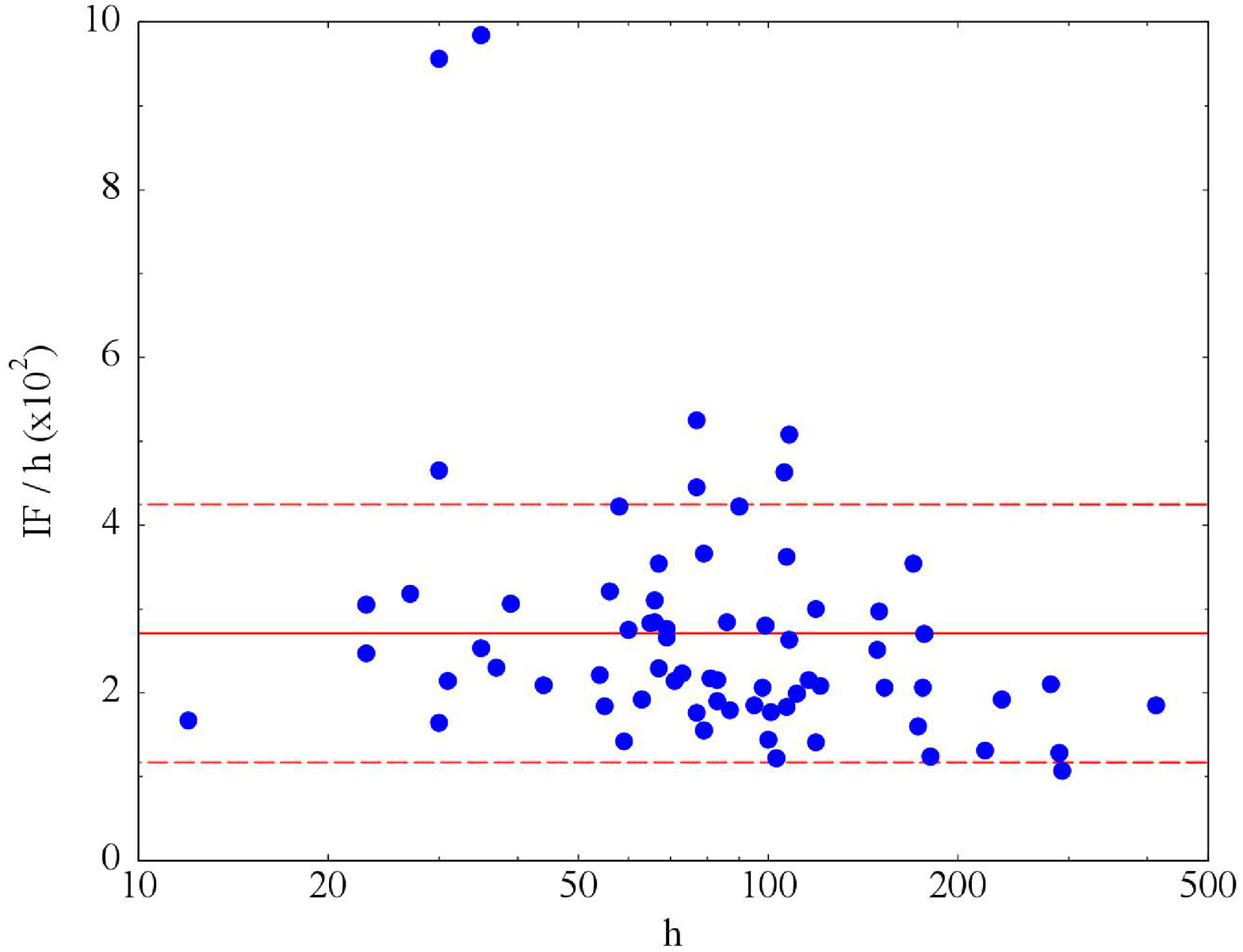}
  \caption{\label{f4} \footnotesize Ratio $\mathrm{IF}/h$  versus $h$ for the 70 physical journals listed in Table~\ref{t1}. The full line gives the average value of $\mathrm{IF}/h \simeq 0.027$ , while the two dashed line differ from the latter by one standard deviation  $\sigma \simeq 0.015$. (from ISI Web of Science).}
\end{figure}

 Table~\ref{t1} shows the values of the $h$-index obtained considering all the articles published between 2001 and 2015 (hence over a period of r = 15 years) by 70 journals, covering most of the literature in physics and ranked according to decreasing IF~\footnote{Since we consider only ``proper'' articles, I did not include in the collection those journal mostly publishing review. I also excluded journals belonging to the Nature group, which show much higher values of the ratio $\mathrm{IF}/h$.}. The value of the linear correlation coefficient between the $2^\mathrm{nd}$  and the $5^\mathrm{th}$ column of the Table ($\simeq 0.73$) shows that there is, as expected, a non negligible  degree of correlation between a journal $h$-index and its IF. Yet, Fig.~\ref{f4} shows that the degree of correlation is just moderate, since the relative standard deviation of the ratio  $\mathrm{IF}/h$  is larger than 0.5. Besides, several journals show an IF which is consistently larger than expected from their $h$-index.
\begin{figure}[t]
\centering
  \includegraphics[width=0.8\columnwidth]{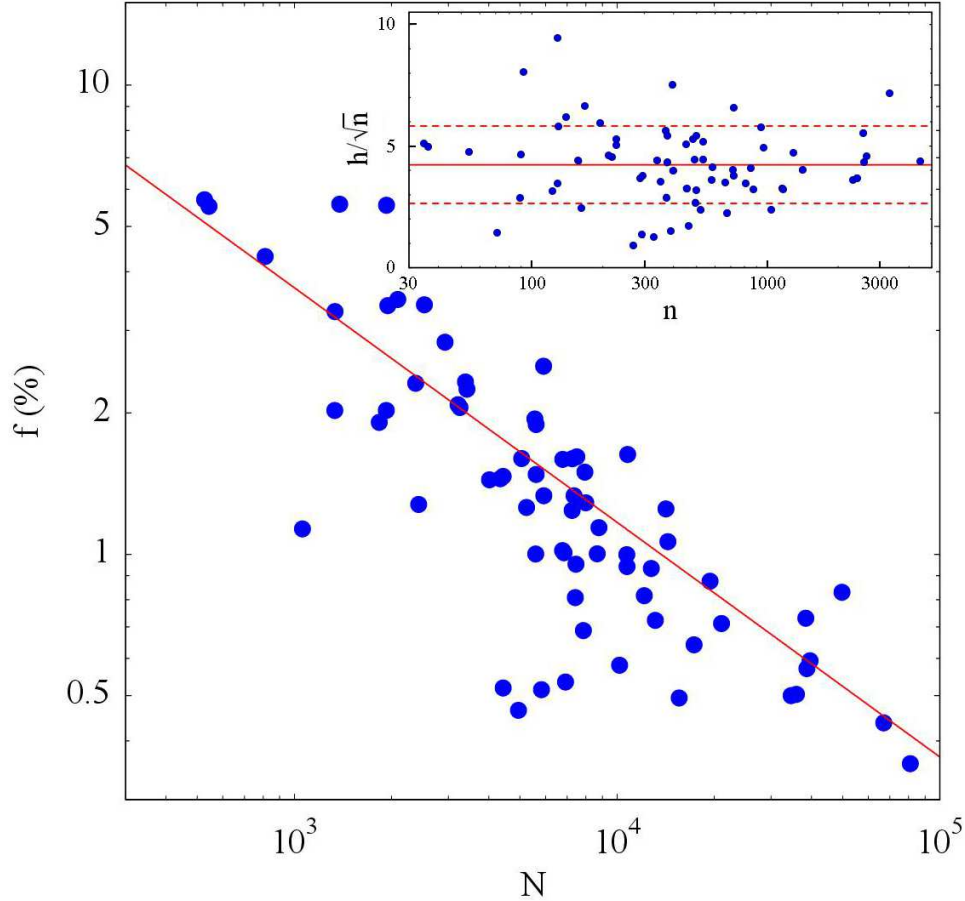}
  \caption{\label{f5} \footnotesize Body: ratio $f = h/N$ for the 70 journals in Table~\ref{t1} versus the total number $N$ of articles published in the period 2001--2015(from ISI Web of Science. The full line is a power law fit $f=aN^{-1/2}$. Inset: ratio $h/\sqrt{N}$ versus the average number of publications $n$ per year, with its average value (full line) and dispersion range to $\pm \sigma$ (broken lines).}
\end{figure}

 Rather surprisingly, a much stronger correlation is found between a journal $h$-index (or, similarly, the fraction $f\le h/N$ of the papers which are cited at least $h$ times) and the total number of publication in the considered period. In fact, the body of Fig.~\ref{f5} shows that, for a set of data covering more than two decades in $N$, $f$ scales approximately as $N^{-1/2}$, which amounts to state that, as shown in the inset, the ratio $h/\sqrt{n}$, where $ n=N/r$ is the average number of publication per year of a journal, is approximately constant.
 This strongly suggests that the main factor determining the ``performance" of a journal, evaluated in terms of its $h$-index, is just \emph{how many papers it publishes}, but that, at the same time, $h$ grows only as $\sqrt{n}$. This evidence bears a curious analogy with a result I previously found~\cite{Piazza2015} for the $h$-index of a moderately large collection of individual physicists, which also seems to scale as the square root of the number of papers they have published in their career~\footnote{More precisely, by examining the individual citation reports of 470 authors of PRL, I found that their $h$-indexes (spanning on the overall the range $h\in [7-60]$) is quite well fit by the relation  $h(n_p) \simeq 2.72 n_p^{1/2} + 2.5$, where $n_p$ is the total number of paper they have published. This formula allows to predict the $h$-index of a colleague to within 20\% when his/her $n_p$ is known.}.

Summarizing, the journal $h$-index seems to be a stable bibliometric parameter that may capture better than the IF the performance of a scientific journal in publishing papers that are read and cited for a long time. Yet, this parameter seems to be basically determined by the total number of papers published by the journal itself. From the publisher's point of view, therefore, it does not seem to be rewarding to split a single subject over several journals covering specific subfields. Giving that, after all, some correlation does exist between a journal IF and $h$-index, this number  may also increase the short-time impact of the journal publications.  To increase their chance to be cited, the authors may also wish to publish in a ``thicker'' journal, provided of course that they keep in mind that the fraction of the papers published by that journal expected to get, in a given number of years, a number of citations $c\ge h$, \emph{decreases} approximately as $\sqrt{N}$ (which of course lowers their chance to enter this ``successful collection").

%

\end{document}